\documentclass[pre,twocolumn,showpacs,preprintnumbers,amsmath,amssymb,superscriptaddress]{revtex4}

\usepackage{graphicx}
\usepackage{dcolumn}
\usepackage{bm}
\usepackage{psfrag}

\providecommand*{\dd}{\mathrm{d}}
\providecommand*{\dod}[2]{\frac{\dd #1}{\dd #2}}
\providecommand*{\op}[1]{\hat{#1}}
\providecommand*{\X}{\op\sigma_x}
\providecommand*{\Y}{\op\sigma_y}
\providecommand*{\Z}{\op\sigma_z}

\providecommand*{\Hop}{\op H}
\providecommand*{\Tr}{\operatorname{Tr}}
\providecommand*{\Kom}[2]{[#1,#2]}
\providecommand*{\Ket}[1]{\left|#1\right>}
\providecommand*{\Bra}[1]{\left<#1\right|}
\providecommand*{\vek}[1]{\bm #1}
\providecommand*{\ie}{\text{i}}

\begin{document}


\title{Modeling heat transport through completely positive maps}

\author{Hannu Wichterich}
\email{hwichter@uni-osnabrueck.de} \affiliation{Fachbereich Physik, Universit\"at Osnabr\"uck, Barbarastrasse 7, D-49069 Osnabr\"uck, Germany}
\author{Markus J. Henrich}
\email{henrich@theo1.physik.uni-stuttgart.de}
\affiliation{Institut f\"ur Theoretische Physik I, Universit\"at 
Stuttgart, Pfaffenwaldring 57, D-70550 Stuttgart, Germany}
\author{Heinz-Peter Breuer}
\affiliation{Physikalisches Institut, Universit\"at Freiburg,
Hermann-Herder-Strasse 3, D-79104 Freiburg, Germany}
\author{Jochen Gemmer}
\affiliation{Fachbereich Physik, Universit\"at Osnabr\"uck, Barbarastrasse 7,
D-49069 Osnabr\"uck, Germany}

\author{Mathias Michel}
\affiliation{%
Advanced Technology Institute,
School of Electronics and Physical Sciences, University of Surrey, Guildford,
GU2 7XH, United Kingdom
}%

\date{\today}

\begin{abstract}
We investigate heat transport in a spin-$\frac{1}{2}$ Heisenberg
chain, coupled locally to independent thermal baths of different
temperature. The analysis is carried out within the framework of
the theory of open systems by means of appropriate quantum master
equations. The standard microscopic derivation of the
weak-coupling Lindblad equation in the secular approximation is
considered, and shown to be inadequate for the description of
stationary nonequilibrium properties like a non-vanishing energy
current. Furthermore, we derive an alternative master equation
that is capable to describe a stationary energy current and, at
the same time, leads to a completely positive dynamical map. This
paves the way for efficient numerical investigations of heat
transport in larger systems based on Monte Carlo wave function
techniques.
\end{abstract}

\maketitle

\section{Introduction}\label{sec:level1}

The transport behavior of one-dimensional systems has intensively
been investigated both in classical and in quantum mechanical
context for several decades now. In the classical domain it seems
to be largely accepted that normal energy transport, i.e., spatial
diffusion instead of ballistic transport or localization requires
the chaotic dynamics of a non-integrable system \cite{Lepri2003,Vollmer2002}. In the
non-classical regime the question whether normal transport
behavior may arise from the underlying quantum mechanical
equations of motion is a controversial issue \cite{zotos-1997,Jung2006,Michel2005}. This is mostly due
to the nontrivial character of the question, how energy or heat is
transported through a system on a microscopic level of
description.

Among the most prominent techniques of theoretical description is
the use of the Green-Kubo Formula \cite{Kubo1957,Heidrich2002} that is derived
on the basis of linear response theory. Here, similar to the
classical case, strong evidence arises that integrability leads to
diverging transport coefficients and that thus normal transport is
not to be expected \cite{zotos-1997}. An overview of theoretical
approaches along with their respective subtleties, pitfalls and
shortcomings was recently given in \cite{michel-2006-20}.

The article at hand addresses the problem within the framework of
the theory of open quantum systems \cite{breuer,Weiss1999,Scully1997} 
by coupling the system of interest explicitly to environments of different 
temperature.
In recent years
many investigations on heat transport have been carried out in
this framework \cite{Saito2000,Saito2003,Michel2003}. According to the results
of these studies numerical evidence for normal transport behavior arises, even in the integrable
system. However, the investigations are generally limited to small system sizes as the computation is
highly demanding.

To investigate a real bulk property of the material, 
e.g., the heat conductivity, one primary aim is to apply methods
that can cope with larger systems in order to rule out that the
mentioned findings are merely due to finite size effects. The
present article constitutes one step towards this aim.

The common setup in models of open quantum systems comprises the
system of interest and an environment. Employing various
approximation schemes for the system-environment coupling one
derives effective dissipative equations of motion for the reduced
density matrix $\op{\rho}_{\text S}$ of the open system. 
These equations are called Markovian quantum master equations (QME). 
Here, the term Markovian refers to the special time-local structure that
arises when all memory effects are negligible. The QME is usually required to be in
Lindblad form in order to guarantee the preservation of the
normalization and of the positivity of the reduced density matrix,
as well as the complete positivity of the resulting dynamical map
\cite{Gorini1976,Lindblad1976}.

Once equipped with a QME of Lindblad structure the way is paved
for efficient numerical studies of larger systems by means of the
stochastic simulation techniques provided by the Monte Carlo wave
function method (see, e.g., \cite{breuer,plenio-knight-1998} and
references therein). 

In contrast to the common open system setup described above,
where the environment serves as a heat bath driving the system into thermal
equilibrium, the introduction of a second heat bath of different temperature leads to
dynamics that may feature a nonequilibrium steady state.
Such states typically feature both energy currents as well as temperature gradients.
Unfortunately, in the present case the derivation of a QME in Lindblad form,
that reasonably describes the expected physical behavior turns out
to be other than straightforward.
Lacking any QME in Lindblad form
also the mentioned efficient stochastic simulation technique is no
longer applicable.
As an example let us
study energy transport in
one-dimensional Heisenberg spin-$\frac{1}{2}$ chains that are
coupled locally to heat baths of different temperature.

In order to illuminate the problem that arises in the attempt to
apply the standard recipe to transport dynamics we shall focus on
one of the aforementioned approximations that are invoked in the derivation of a QME which actually does feature Lindblad form. The most common approximation which yields a Lindblad form (in the weak-coupling limit) is the so-called secular approximation (SA) \cite{Davies1974b,Spohn}. Essentially, this approximation consists
in replacing the generator of the interaction picture master
equation by its time average. The SA is justified as long as the
time scale set by the differences of the Bohr frequencies of the
open system is short compared to the relaxation time. In many
cases it is equivalent to the rotating wave approximation, but
there exist counterexamples \cite{maniscalco-2004-6}.

We observe that, once the SA has been performed, one is left with a QME that
predicts a vanishing energy current in the presence of a finite
temperature gradient. The standard weak-coupling Lindblad master
equation that is obtained on the basis of the SA is therefore
inappropriate to describe heat transport. Hence, in order to be
able to apply the standard Monte Carlo wave function techniques to
heat transport one needs an appropriate quantum master equation
which leads to a finite stationary energy current and, at the same
time, is of Lindblad form. Here, we derive such a master equation
which is valid in the regime of weak internal couplings.

The paper is organized as follows. In Sec.~\ref{BACKGROUND} we
introduce the details of our model and give the definition of the
energy current operator. Moreover, we briefly recall the
derivation of the Redfield master equation for the density matrix
of the spin chain. Section \ref{sec:level4} contains a discussion
of the SA and of its influence in the description of heat
transport. We further derive a completely positive Markovian
master equation for the dynamics of the spin chain and demonstrate
that it yields an excellent approximation of the dynamics given by
the Redfield equation. Several conclusions and implications for
future investigations are presented in Sec.~\ref{CONCLU}.

%
%
\section{Theoretical Background}
\label{BACKGROUND}

%
%
\subsection{The Model and the current operator}
\label{sec:level21}

\begin{figure}[hbt]
\includegraphics[width=0.43\textwidth]{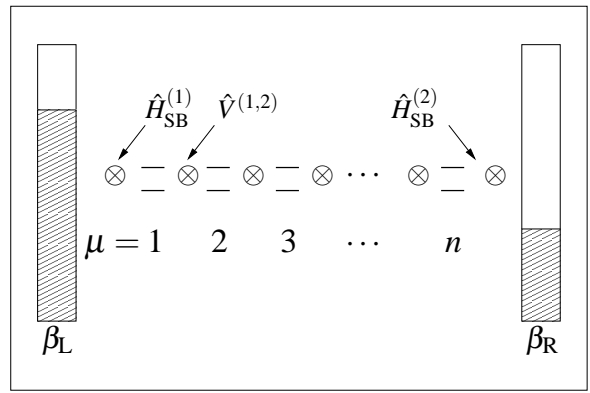}
\caption{Spin-$\frac12$ chain coupled to heat baths of different temperature. The stationary state features a non-vanishing energy flux through the chain of spins from the hotter towards the colder heat bath.}
\label{fig:spinkette}
\end{figure}

We investigate heat transport in spin-$\frac12$ chains of length $n$, coupled to heat baths of different temperature at both ends (see Fig.~\ref{fig:spinkette}).
The Hamiltonian of the spin chain reads
\begin{equation}\label{hamiltonian}
	\op{H}_{\text{S}}=\op{H}_{\text{loc}}+\op{V}\,.
\end{equation}
The first term describes a local energy contribution due to an external field of strength $\Omega$,
\begin{equation}
	\op{H}_{\text{loc}}=\sum\limits_{\mu=1}^{n}\frac{\Omega}{2}\,\Z^{(\mu)}\,,
\end{equation}
whereas the second term accounts for a homogeneous nearest-neighbor Heisenberg interaction with a coupling parameter $\lambda$,
\begin{equation}
	\op{V}=\sum_{\mu}\,\op{V}^{(\mu,\mu+1)}
	=\lambda\sum_{\mu}\,\op{\vek{\sigma}}^{(\mu)}\cdot\op{\vek{\sigma}}^{(\mu+1)}\,.
\end{equation}
$\op{\vek{\sigma}}\equiv(\X,\Y,\Z)^T$ is the spin vector operator with the well-known Pauli spin
matrices as its elements. 
Braced upper indices label the respective spin.

In order to obtain an operator for the energy current between two adjacent spins in our system, we consider the time evolution of the local energy operator given by the Heisenberg equation of motion for operators at site $\mu$ 
\begin{equation}\label{heisenbergeqnofmotion}
	\dod{\op{H}_{\text{loc}}^{(\mu)}}{t}
	=\ie\Kom{\op{H}_{\text{S}}}{\op{H}_{\text{loc}}^{(\mu)}}
	+\frac{\partial}{\partial t}\op{H}_{\text{loc}}^{(\mu)}\,.
\end{equation}
Since $\op{H}_{\text{loc}}^{(\mu)}$ is not explicitly time dependent Eq.~(\ref{heisenbergeqnofmotion}) becomes after inserting Eq.~(\ref{hamiltonian})
\begin{equation}\label{currents}
	\dod{\op{H}_{\text{loc}}^{(\mu)}}{t}
	=\ie\left(\Kom{\op{V}^{(\mu-1,\mu)}}{\op{H}_{\text{loc}}^{(\mu)}}
	+\Kom{\op{V}^{(\mu,\mu+1)}}{\op{H}_{\text{loc}}^{(\mu)}}\right)\,.
\end{equation}
Assuming that the local energy is a conserved quantity, which is justified when $\Omega$ is large compared to $\lambda$, we can rewrite Eq.~(\ref{currents}) as
\begin{equation}\label{continuityeqn}
	\dod{\op{H}_{\text{loc}}^{(\mu)}}{t}
	=\mathrm{div}\op{J}=\op{J}^{(\mu,\mu+1)}-\op{J}^{(\mu-1,\mu)}\,,
\end{equation}
where we introduced a discretized version of the continuity equation. By comparing Eqs.~(\ref{currents}) and
(\ref{continuityeqn}) we deduce
\begin{equation}\label{currentop}
	\op{J}^{(\mu,\mu+1)}
	=\ie\,\Kom{\op{V}^{(\mu,\mu+1)}}{\op{H}_{\text{loc}}^{(\mu)}}\,,
\end{equation}
being the energy current flowing from site $\mu$ to site $\mu + 1$. 

\subsection{Redfield Equation}
\label{sec:level22}

The total system, i.e., system and both heat baths is described by the Hamiltonian
\begin{equation}\label{htot}
	\op{H}_{\text{tot}} 
	= \op{H}_{\text{S}} 
	+ \sum_{i=1}^2 \left( \op{H}_{\text{B}}^{(i)} 
	+ \op{H}_{\text{SB}}^{(i)} \right)\,.
\end{equation}
$\op{H}_{\text{B}}^{(i)}$ stands for the Hamiltonian of the $i$th bath, $\op{H}_{\text{SB}}^{(i)}$ denotes the respective system-bath interaction.

In the following we will first turn to a scenario with a single bath for the purpose of a compact notation.
The time evolution of the total system is described by the von Neumann equation which reads
\begin{equation}\label{lvn}
	\dod{\op{\rho}_{\text{tot}}}{t}
	=-\ie\Kom{\op{H}_{\text{tot}}}{\op{\rho}_{\text{tot}}}\,.
\end{equation}
$\op{\rho}_{\text{tot}}$ represents the density operator of the total system.
The aim is to derive an equation of motion for the density operator $\op{\rho}_{\text S}$ of the spin chain which is defined through the partial trace taken over the bath variables,
\begin{equation}\label{partialtrace}
	\op{\rho}_{\text S}(t)
	=\Tr_{\text{B}}\left\lbrace \op{\rho}_{\text{tot}}(t)\right\rbrace\,.
\end{equation}
To this end, we invoke the Born-Markov approximation and assume that the bath is in a thermal equilibrium state at inverse temperature $\beta$. 

If the interaction Hamiltonian is cast into the form
\begin{equation}\label{sysbathcouplinghamiltonian}
	\op{H}_{\text{SB}} = \sum_k\op{X}_k\otimes\op{Y}_k
\end{equation}
with Hermitian operators $\op{X}_k$ and $\op{Y}_k$ of system and bath respectively, the standard procedure (cf.~\cite{breuer}) yields the following Schr\"odinger picture QME
\begin{equation}\label{rme}
	\dod{\op{\rho}_{\text S}}{t} 
	=-\ie\Kom{\op{H}_{\text S}}{\op{\rho}_{\text S}}
	+\mathcal{D}(\op{\rho}_{\text S})\,.
\end{equation}
The first term describes the free dynamics, whereas the dissipative part
\begin{align}\label{dissipator}
	\mathcal{D}(\op{\rho}_{\text S})
	=\sum\limits_{kl}\int_0^{\infty}\dd\tau\,
	\Gamma_{lk}(\tau,\beta)\,\Kom{\op{X}_k(-\tau)\,
	\op{\rho}_{\text S}}{\op{X}_l} + \mathrm{H.c.}
\end{align}
accounts for the influence of the environment.
The time dependence of the operator $\op{X}_k(-\tau)$ in Eq.~(\ref{dissipator}) has to be interpreted as
\begin{equation}\label{eq:14}
	\op{X}_k(-\tau) = e^{-\ie\op{H}_{\text S}\tau} \op{X}_k e^{\ie\op{H}_{\text S}\tau}\,.
\end{equation}
The bath inverse temperature $\beta$ is embodied in the bath correlation functions
\begin{align}\label{bathcorrelations}
	\Gamma_{kl}(\tau,\beta)&=\langle\op{Y}_k(\tau)\op{Y}_l\rangle_{\text B}
	=\Tr_{\text B}\lbrace\op{\rho}_{\text B}\,\op{Y}_k(\tau)\op{Y}_l \rbrace
\end{align}
with
\begin{align}
	\op{\rho}_{\text B}
	&=\frac{e^{-\beta\op{H}_{\text B}}}{\Tr_{\text B}\lbrace 
	e^{-\beta\op{H}_{\text B}}\rbrace}\,.
\end{align}
Equation (\ref{rme}) is known as Redfield master equation, and has many applications ranging from NMR to the description of chemical dynamical systems \cite{may,pollard}.
We will use this equation in the following as a reference point to judge whether further approximations are suitable or not.

%
%
\subsection{Local Environment Coupling}
\label{verueckte_Idee_von_Mathias}

The chain of $n$ spins shall be coupled only locally to the respective baths, i.e., via the outermost spins (cf.~Fig.~\ref{fig:spinkette}).
Consider, e.g., the system operator that couples to the left heat bath
\begin{equation}\label{localoperator}
	\op{X}_{\text L} = \X^{(1)}\otimes\op 1^{(2)}\otimes\cdots\otimes\op 1^{(n)}
	\equiv\X^{(1)}\,.
\end{equation}
Thus, the double sum in Eq.~(\ref{dissipator}) contains only one single term. 
The corresponding Redfield dissipator reads
\begin{equation}\label{leftdissipator}
	\mathcal{D}_{\text L} (\op{\rho}_{\text S})
	=\int_0^{\infty}\dd\tau\,\Gamma(\tau,\beta_{\text L})
	\Kom{\op{X}_{\text L}(-\tau)\,\op{\rho}_{\text S}}{\op{X}_{\text L}} + \mathrm{H.c.}
\end{equation}
We choose a simple bath modelled by an infinite number of independent harmonic oscillators,
\begin{equation}\label{bathoscillators}
	\op{H}_{\text{B}}
	=\sum\limits_{k=1}^{\infty}\omega_k\,\op{b}_k^{\dagger}\op{b}_k\,,
\end{equation}
with the usual bosonic creation an annihilation operators $\op{b}_k^{\dagger},\,\op{b}_k$.
The system-environment coupling as defined in Eq.~(\ref{sysbathcouplinghamiltonian}) is taken to be linear in the oscillator amplitudes,
\begin{equation}\label{bathoscillatoramplitudes}
	\op{Y}_{\text L}=\sum\limits_{k=1}^{\infty}c_k
	\op{b}_k^{\dagger}+c_k^{\ast} \op{b}_k\,,
\end{equation}
where the $c_k$ are coupling constants. Plugging Eq.~(\ref{bathoscillatoramplitudes}) into Eq.~(\ref{bathcorrelations}) the bath correlation function is found to be
\begin{align}\label{oszibathcorrelationstimedomain}
	\Gamma(\tau,\beta_{\text L})&=\int_{-\infty}^{\infty}\dd\omega\,
	e^{\ie\omega\tau}\,\Gamma(\omega,\beta_{\text L})
\end{align}
with
\begin{align}\label{oszibathcorrelations}
	\Gamma(\omega,\beta_{\text L})&=\frac{\kappa}{2}\,
	\left[J(\omega)-J(-\omega)\right]\, N(\omega,\beta_{\text L})\,.
\end{align}
Here, $J(\omega)$ denotes the spectral density, $N(\omega,\beta_{\text L})$ represents the Planck distribution
\begin{equation}\label{boseeinstein}
	N(\omega,\beta_{\text L})=\frac{1}{e^{\beta_{\text L}\omega}-1}\,,
\end{equation}
and $\kappa$ is a coupling parameter. 
Several forms of the spectral density are considered in the literature, we choose an Ohmic bath, i.e.,
\begin{equation}\label{ohmic}
	J(\omega)=\Theta(\omega)\,\omega\,,
\end{equation}
where $\Theta(\omega)$ is the Heaviside step function defined by
\begin{equation}\label{heaviside}
	\Theta(\omega)=
	\begin{cases}
	1 & \;\omega > 0\,, \\
	0 & \;\omega \leq 0\,.
	\end{cases}
\end{equation}

For a proper heat conduction model a second bath of different temperature is coupled analogously to the opposite side of the chain.
The final form of the QME thus reads
\begin{equation}\label{twobathscenario}
	\dod{\op{\rho}_{\text S}}{t} =-\ie\Kom{\op{H}_{\text S}}{\op{\rho}_{\text S}}
	+\mathcal{D}_{\text L}(\op{\rho}_{\text S})
	+\mathcal{D}_{\text R}(\op{\rho}_{\text S})\,.
\end{equation}

A full numerical investigation based on such a type of master equation may be found in~\cite{Saito2000,Saito2003}.
More generally the use of QME's of Redfield type like Eq.~(\ref{rme}), respectively Eq.~(\ref{twobathscenario}) is common practise \cite{may,pollard}, regardless of the well-known disadvantage, that these forms may violate the preservation of positivity of the reduced density operator.

This violation is usually observed at the very early stage of the relaxation process only and may possibly be cured by a slippage of initial conditions \cite{Suarez1992,Gnutzmann1996}. Moreover - considering an equilibrium scenario ($\beta_L=\beta_R\equiv\beta$) for a moment -  the canonical Gibbs state 
\begin{align}
\op{\rho}^{\mathrm G}_{\text S}\equiv\frac{e^{-\beta\op{H}_{\text S}}}{\Tr\lbrace{e^{-\beta\op{H}_{\text S}}}\rbrace}
\end{align}
is a stationary solution of the master equation (\ref{twobathscenario}) \cite{Kubo1991}. Whether the stationary solution of Eq.~(\ref{twobathscenario}) is likewise well defined and necessarily positive for $\beta_L\neq\beta_R$ requires further investigation.
In view of these remarks we emphasize that we do not intend here to depart from Eq.~(\ref{twobathscenario}) in favor of preservation of positivity itself, but because Eq.~(\ref{twobathscenario}) can not be treated with the standard Monte Carlo wave function technique.
Although extended stochastic schemes exist for the solution of general QME's as the Redfield equation~\cite{Breuer2}, these methods turn out to be less efficient, in general, than the standard approach for a Lindblad QME (see Section~\ref{sec:level4}).
Therefore, it would be highly desireable to derive a Lindblad type of master equation for the heat conduction scenario given in Fig.~\ref{fig:spinkette}.
%
%
\section{Complete Positivity vs.\ Nonequilibrium Steady States}
\label{sec:level4}

In an axiomatic approach to the theory of quantum dynamical semigroups the generator $\mathcal{L}$ of the master equation $\dot{\op{\rho}}_{\text S}=\mathcal{L}\op{\rho}_{\text S}$ can be shown to have the most general form \cite{Gorini1976,Lindblad1976}
\begin{equation}\label{lindbladform}
	\mathcal{L}\op{\rho}_{\text S}=
	-\ie\Kom{\op{H}_{\text S}}{\op{\rho}_{\text S}}
	+\sum_k \alpha_{k}
	\Big( \op{L}_k \op{\rho}_{\text S} \op{L}^{\dagger}_k
	- \frac{1}{2} [\op{L}^{\dagger}_k \op{L}_k, \op{\rho}_{\text S}]_{+}\Big)\,,
\end{equation}
where $[\dots,\dots]_+$ denotes the anti-commutator and the $\op{L}_k$ are operators acting on the Hilbert space of the open system, and the $\alpha_k$ are non-negative numbers.
A generator of this form is said to be a Lindblad generator.
It guarantees the preservation of the trace and of the positivity of the density matrix.
Moreover, the dynamical semigroup generated by ${\mathcal{L}}$ obeys the property of complete positivity \cite{breuer,Alicki}.

In general, the Redfield master equation (\ref{rme}), respectively (\ref{twobathscenario}), is in Lindblad form only if further approximations are carried out.
On the one hand, in many applications the violation of complete positivity is reasonably small and might not even show up for a large class of initial conditions \cite{Gnutzmann1996}.
On the other hand, it is always advantageous to have a Lindblad equation at hand, especially for numerical purposes.
In fact, the dynamics of any QME with a generator of the form of Eq.~(\ref{lindbladform}) can be represented through a stochastic process in Hilbert space.
This representation gives rise to the Monte Carlo wave function or quantum trajectory method \cite{breuer,plenio-knight-1998}, which constitutes a promising technique to cope with open quantum systems having many degrees of freedom.
In the present Section we consider two different types of derivations that aim at a QME of Lindblad form (\ref{lindbladform}) for our model of heat conduction (see Fig.~\ref{fig:spinkette}).

%
%
\subsection{Secular Approximation}
\label{sec:level4a}

The secular approximation can be conveniently studied by transforming to the interaction picture.
In this representation the Redfield master equation reads
\begin{equation}\label{eq9}
	\dod{\op{\rho}^{\text I}_{\text S}}{t}=\mathcal{D}(\op{\rho}_{\text S}^{\text I})
\end{equation}
with the Dissipator
\begin{equation}\label{eq9b}
	\mathcal{D}(\op{\rho}_{\text S}^{\text I})
	=\sum_{kl}\int_0^{\infty}\dd\tau\,\Gamma_{lk}(\tau,\beta)
	\Kom{\op{X}_k(t-\tau)\,\op{\rho}^{\text I}_{\text S}}{\op{X}_l(t)} + \mathrm{H.c.}
\end{equation}

In order to carry out the secular approximation we choose a suitable form for the operators $\op X_k(t)$ in the interaction picture.
Thus -- following \cite{breuer} -- we make use of a decomposition of these operators into eigenoperators of
$\Hop_\text{S}$. 
With the help of projection operators $\op\Pi(\epsilon)$ that project onto the eigenspaces associated with
the eigenvalues $\epsilon$ of $\Hop_\text{S}$ we define
\begin{equation}\label{eq10}
	\op X_k(\omega)=\sum_{\epsilon-\epsilon'=\omega} \op \Pi (\epsilon) \op X_k \op
	\Pi(\epsilon')\,,
\end{equation}
where $\omega$ represents a certain energy difference (Bohr frequency) of the system.
Summing over all possible frequencies yields the desired eigenoperator decomposition
\begin{equation}\label{eq10b}
	\op X_k=\sum_\omega \op X_k(\omega)\,.
\end{equation}
The operators (\ref{eq10}) obey the relation
\begin{equation}\label{eq11}
	\op X^{\dagger}_k(\omega)=\op X_k(-\omega)\,.
\end{equation}
The interaction picture operators are now easily obtained from the relations
\begin{eqnarray}\label{eq12}
	e^{\ie \Hop_\text{S} t}\, \op X_k(\omega)\, e^{-\ie \Hop_\text{S} t}
	& = &  e^{-\ie \omega t}\,\op X_k(\omega)\,, \\\label{eq13}
	e^{\ie \Hop_\text{S} t}\, \op X_k^\dagger(\omega)\, e^{-\ie \Hop_\text{S} t}
	& = & e^{\ie \omega t}\, \op X_k^\dagger(\omega)\,.
\end{eqnarray}
In the case of a single system operator [see Eq.~(\ref{localoperator})] the indices $k,l$ in Eq.~(\ref{eq9b})
may be dropped.
Inserting Eqs.~(\ref{oszibathcorrelationstimedomain}), (\ref{eq12}) and (\ref{eq13}) into Eq.~(\ref{eq9b}) and making use of the formula
\begin{equation}\label{halfsidedfouriertransform}
	\int_{0}^{\infty}\dd\tau\,
	e^{\ie(\omega-\Omega)\tau}=\pi\,\delta(\omega-\Omega)+\ie\frac{P}{\omega-\Omega},
\end{equation}
where $P$ indicates the Cauchy principal value.
The dissipator takes the form
\begin{align}\label{eq16}
	\mathcal{D}_\text{L}(\op{\rho}_{\text S}^{\text I})
	=& \pi \sum_{\omega,\omega'}
	e^{\ie(\omega'-\omega)t} \Gamma_\text{L}(\omega) \times \nonumber \\
	&\quad\times \Kom{\op{X}_\text{L}(\omega)\,\op{\rho}^{\text I}_{\text S}}
	{\op{X}_\text{L}^\dagger(\omega')}+\mathrm{H.c.}
\end{align}
The imaginary part of (\ref{halfsidedfouriertransform}) merely leads to a small energy shift and will silently be incorporated in the Hamiltonian~(\ref{hamiltonian}) henceforth. 
By performing the secular approximation one neglects the rapidly oscillating terms in Eq.~(\ref{eq16}). 
Thus all terms with $\omega' \ne \omega$ will be neglected
\begin{align}\label{eq17b}
	\dod{\op{\rho}^{\text I}_{\text S}}{t} 
	=\pi \sum_{\omega} \Gamma_{\text{L}}(\omega)\Big(
	&\Kom{\op{X}_{\text{L}}(\omega)\,\op{\rho}^{\text I}_{\text S}}{\op{X}_{\text{L}}^\dagger(\omega)}
	\notag\\
	&\quad+\Kom{\op{X}_{\text{L}}(\omega)\,\op{\rho}^{\text I}_{\text S}}
	{\op{X}_{\text{L}}^\dagger(\omega)}^{\dagger}\Big)
\end{align}
Since $\Gamma_\text{L}(\omega)$ is a positive quantity for all $\omega$, Eq.~(\ref{eq17b}) is of Lindblad form (\ref{lindbladform}) as can be readily seen by a slight rearrangement of terms
\begin{align}
	\label{eq:38} 
	\dod{\op{\rho}^{\text I}_{\text S}}{t}
	=2\pi\sum_{\omega}\Gamma_{\text{L}}(\omega)
	\Big(&\op{X}_{\text{L}}(\omega)\op{\rho}^{\text I}_{\text S}\op{X}_{\text{L}}^{\dagger}(\omega)
	\notag\\
	&\quad-\frac{1}{2}\Kom{\op{X}_{\text{L}}^{\dagger}(\omega)\op{X}_{\text{L}}(\omega)}
	{\op{\rho}^{\text I}_{\text S}}_{+}\Big)\,.
\end{align}
For a heat conduction model as shown in Fig.~\ref{fig:spinkette} a second dissipator $\mathcal{D}_\text{R}(\op \rho_\text{S})$ will be added to Eq.~(\ref{eq9}) and we end up with an equation analogous to Eq.~(\ref{twobathscenario}).

For systems with a non-degenerate spectrum like the weakly coupled spin-chains treated in this work, the SA decouples the equations of motion for the populations and coherences, i.e., for the diagonal and for the off-diagonal elements of $\op \rho^{\text I}_{\text S}$ in the eigenbasis of the system Hamiltonian $\op H_{\text S}$ \cite{Alicki,breuer,Stollsteimer2006,Blumel1991}, even and especially in the case of two baths with different temperatures.

This decoupling gives rise to two independent systems of linear differential equations, one for the diagonal and one for the off-diagonal elements of $\op{\rho}_{\text S}$ in the energy representation. If we further assume that the stationary solution of Eq.~(\ref{eq:38}) is unique, then this implies that the off-diagonal elements of $\op{\rho}_{\text S}$ decay and the stationary state $\op{\rho}_{\text S}^{\text{st}}$ is diagonal in the energy representation.

The stationary state can therefore be decomposed as
 \begin{align}
\op{\rho}_{\text S}^{\text{st}}&=\sum\limits_{n=1}^{d}P_n^{\text{st}}\;\Ket{n}\Bra{n},\\
\op{H}_{\text S}\Ket{n}&=\epsilon_n\Ket{n}
\end{align}
where $d$ denotes the dimension of the Hilbert space under consideration and $\Ket{n}$ is the eigenvector corresponding to the eigenvalue $\epsilon_n$ of $\op{H}_{\text S}$. 
The expectation value of an operator $\op J$ in the stationary state then takes the form
\begin{align}\nonumber
\langle \op J\rangle^{\text{st}} &\equiv\Tr\big\lbrace\op{\rho}_{\text S}^{\text{st}}\,\op J\big\rbrace \\\label{eq:jstat}
&=\sum\limits_{n=1}^{d}P^{\text{st}}_n\,\Bra{n}\op J\Ket{n}~.
\end{align}
We now require $\op J$ to be an energy current operator. A plausible requirement on such an operator is
\begin{align}\label{eq:jopcond}
\Bra{n}\op J\Ket{n}=0~,
\end{align}
since the energy eigenstates $\Ket{n}$ must not carry energy currents, when the system is chain-like with open boundary conditions. It can be shown in a straight forward manner that our particular choice of the current operator (\ref{currentop}) satisfies the condition in Eq. (\ref{eq:jopcond}). Equation (\ref{eq:jopcond}) enters Eq.~(\ref{eq:jstat}) and leads to our final result
\begin{align}\label{eq:vanish}
\langle \op J\rangle^{\text{st}}=0~.
\end{align}

Summarizing, the QME in the SA [cf. Eq.~(\ref{eq:38})] features a steady state solution $\op{\rho}_{\text S}^{\text{st}}$ that is diagonal in the energy representation, and by any plausible choice of a current operator, stationary currents will vanish by virtue of Eq.~(\ref{eq:vanish}). This prediction is most unphysical in view of the nonequilibrium scenario that is depicted in Fig.~\ref{fig:spinkette}~.
 
In this sense the QME in the SA proves inappropriate for our purposes. This failure of the SA is physically plausible since, as already mentioned, it only applies if the differences of the Bohr frequencies are large compared to the relaxation rates of the system. In chainlike systems, however, differences of Bohr frequencies between energy levels  within the bands will go to zero with chainlengths going to infinity (even in the limit of strong couplings). Thus, the differences of the Bohr frequencies will eventually be too small to justify the application of the SA in a nonequilibrium scenario.

%
%
\subsection{Weak Internal Coupling Approach}
\label{sec:weakcoupling}
In this Section we shall derive a Lindblad QME that retains
the property of Eq.~(\ref{twobathscenario}) of featuring a nonequilibrium
steady state with non-vanishing energy current, and is thus
suitable for the description of energy transfer.
To this end, we assume that for weak couplings of neighboring spins $\op{X}(-\tau)$ [cf.~Eq.~(\ref{eq:14})] is approximately given by the free dynamics (see \cite{Kleinekathoefer}), i.e.,
\begin{equation}\label{dda}
	\op{X}(-\tau) = e^{-\ie\op{H}_{\text S}\tau} \op{X}
	e^{\ie\op{H}_{\text S}\tau} \simeq e^{-\ie\op{H}_{\text{loc}}\tau}
	\op{X} e^{\ie\op{H}_{\text{loc}}\tau}\,.
\end{equation}
We would like to stress that the weak coupling assumption must be considered fundamental for the very model by virtue of Eqs.~(\ref{currentop}) and (\ref{twobathscenario}). 
The time dependent version of the operator (\ref{localoperator}) thus takes the simple form
\begin{equation}\label{weakcouplingapprox}
	\op{X}_{\text L}(-\tau) \simeq e^{-\ie\Omega\tau}\op{\sigma}^{(1)}_+
	+ e^{\ie\Omega\tau}\op{\sigma}^{(1)}_-\,.
\end{equation}
This approximation enters Eq.~(\ref{leftdissipator}) and yields
\begin{align}\label{localdissipator}
	\mathcal{D}_{\text L}&(\op{\rho}_{\text S})\simeq
	\int_0^{\infty}\dd\tau\,
	\Gamma(\tau,\beta_{\text L})\,\times\\\nonumber 
	&\times\,\left(
	e^{-\ie\Omega\tau}
	\Kom{\op{\sigma}^{(1)}_+\,\op{\rho}_{\text S}}{\X^{(1)}}
	+e^{\ie\Omega\tau}
	\Kom{\op{\sigma}^{(1)}_-\,\op{\rho}_{\text S}}{\X^{(1)}}\right) 
	+\mathrm{H.c.}
\end{align}
with $\Gamma(\tau,\beta_{\text L})$ given by Eq.~(\ref{oszibathcorrelationstimedomain}).
By making use of Eq.~(\ref{halfsidedfouriertransform}) and neglecting the imaginary part (see remarks in Sec.~\ref{sec:level4a}) we are left with
\begin{align}\label{localdissipator2}
	\mathcal{D}_{\text L}(\op{\rho}_{\text S})\simeq
	&\pi\,\Gamma_{\text L}(\Omega)
	\Kom{\op{\sigma}^{(1)}_{\text +}\,\op{\rho}_{\text S}}{\X^{(1)}}\\\nonumber 
	&\quad+\pi\,
	\Gamma_{\text L}(-\Omega)
	\Kom{\op{\sigma}^{(1)}_-\,\op{\rho}_{\text S}}{\X^{(1)}} 
	+\mathrm{H.c.}
\end{align}

To proceed we introduce a complete set of operators $\op{F}_i$ acting on the Hilbert space of the first spin. 
It is convenient to choose this operator basis to be orthonormal with respect to the Hilbert-Schmidt scalar product
\begin{equation}\label{HSorth}
	(\op{F}_i,\op{F}_j)\equiv\Tr\{\op{F}^{\dagger}_i\op{F}_j\}=\delta_{ij}\,.
\end{equation}
A suitable set of operators satisfying these conditions is given by the set 
$\{\op{\sigma}_{+},\op{\sigma}_{-},\Z /2,\op{1}/2\}$.
With the help of these operators Eq.~(\ref{localdissipator2}) can be cast into the form
\begin{equation}\label{standardform1}
	\mathcal{D}_{\text L} (\op{\rho}_{\text S}) =
	\sum_{k,l=1}^2 \gamma_{kl} 
	\Big( \op{F}_k \op{\rho}_{\text S} \op{F}^{\dagger}_l 
	- \frac{1}{2} 
	[\op{F}^{\dagger}_l \op{F}_k, \op{\rho}_{\text S}]_{+}\Big)
\end{equation}
with
\begin{align}\nonumber
\op{F}_1 &= \op{\sigma}_{+}^{(1)}\otimes\op 1^{(2)}\otimes\cdots\otimes\op 1^{(n)}\,,\\
\op{F}_2 &= \op{\sigma}_{-}^{(1)}\otimes\op 1^{(2)}\otimes\cdots\otimes\op 1^{(n)}\,.
\end{align}
The coefficient matrix $\gamma=(\gamma_{kl})$ introduced in Eq.~(\ref{standardform1}) reads
\begin{align}\label{coeff}
	\gamma = \pi\,\left(
	\begin{array}{cc}
		2\,\Gamma_{\text L}(\Omega)& \Gamma_{\text L}(\Omega)+\Gamma_{\text L}(-\Omega) \\
		\Gamma_{\text L}(\Omega)+\Gamma_{\text L}(-\Omega)& 2\,\Gamma_{\text L}(-\Omega)
\end{array}
\right)\,.
\end{align}
A given dissipator of the form of Eq.~(\ref{standardform1}) is in Lindblad form if and only if the coefficient matrix $\gamma$ is positive (see, e.g., \cite{breuer}).
However, the matrix given by Eq.~(\ref{coeff}) is not positive because $\gamma_{11} > 0$ and
\begin{align}
	\mathrm{det}(\gamma)=-\pi\big[\Gamma_{\text{L}}(\Omega)-\Gamma_{\text{L}}(-\Omega)\big]^2<0
\end{align}
since $\Gamma_{\text{L}}(\pm\Omega)\in\mathbb{R}$.
Hence, we conclude that the dissipator derived above is not in Lindblad form.

We suggest the following strategy to bring the dissipator given by Eq.~(\ref{standardform1}) into Lindblad form by only minimal modifications, without invoking the SA. 
The idea is to separate off the largest possible contribution to $\gamma$ that leads to a generator in Lindblad form. 
To this end, we first decompose the coefficient matrix as
\begin{equation}
	\gamma \equiv \gamma_A + \gamma_B\,,
\end{equation}
where
\begin{align}	
	\gamma_A &=\pi\left(
	\begin{array}{cc}
		\gamma_{11} & M \\
		M & \gamma_{22}
	\end{array}
	\right)\,,\\
	\gamma_B &=\pi\left(
	\begin{array}{cc}
		0 & \gamma_{12}- M\\
		\gamma_{21} - M & 0
	\end{array}
	\right)\,.
\end{align}
Of course, such a decomposition holds true for any value of the parameter $M$.
We fix this parameter by the requirement that the determinant of the matrix $\gamma_A$ vanishes,
\begin{equation}\label{COND}
	\mathrm{det}\left(\gamma_A\right) \stackrel{!}{=} 0\,.
\end{equation}
This condition implies that $\gamma_A$ is positive, having one zero eigenvalue. 
Obviously, the condition (\ref{COND}) is satisfied if we choose
\begin{equation}\label{Msolution}
	M=2\sqrt{\Gamma_L(\Omega)\Gamma_L(-\Omega)}\,.
\end{equation}
With this choice the matrix $\gamma_B$ becomes
\begin{equation}
	\gamma_B =\pi\kappa\,J(\Omega)\;
	\left(
	\begin{array}{cc}
		0 & A\\
		A & 0
	\end{array}
	\right)\,,
\end{equation}
where
\begin{equation}
	A \equiv N+\frac12-\sqrt{N^2+N}
\end{equation}
with the Planck distribution $N\equiv N(\beta,\Omega)$ [cf.~Eq.~(\ref{boseeinstein})]. 
We see that the non-zero elements of $\gamma_B$ rapidly vanish as the temperature and, likewise, $N$ goes
to infinity
\begin{equation}
	\lim_{T\rightarrow\infty}A=0\,.
\end{equation}
For not too low temperatures the quantity $A$ is small, and we may neglect the contribution from $\gamma_B$ to the dissipator, ending up, finally, with a master equation in Lindblad form.
Neglecting $\gamma_B$ in an appropriate parameter range can be considered as a minor invasion in comparison to the numerous assumptions that lead to Eq.~(\ref{standardform1}).
Summarizing, the final form of the QME in the weak internal coupling limit thus reads
\begin{equation}\label{DDAqme}
	\dod{\op{\rho}_{\text S}}{t} 
	=-\ie\Kom{\op{H}_{\text S}}{\op{\rho}_{\text S}}
	+\mathcal{D}_{\text L}(\op{\rho}_{\text S})
	+\mathcal{D}_{\text R}(\op{\rho}_{\text S})\,,
\end{equation}
where the dissipator for the left heat bath is given by
\begin{equation}
	\mathcal{D}_{\text L} (\op{\rho}_{\text S}) 
	= \sum_{k,l=1}^2 (\gamma_A)_{kl} 
	\Big( \op{F}_k \op{\rho}_{\text S} \op{F}^{\dagger}_l 
	- \frac{1}{2}
	[\op{F}^{\dagger}_l \op{F}_k, \op{\rho}_{\text S}]_{+}\Big),
\end{equation}
and the dissipator $\mathcal{D}_{\text R}(\op{\rho}_{\text S})$ for the right heat bath is defined correspondingly.

\begin{figure}
\includegraphics[width=0.45\textwidth]{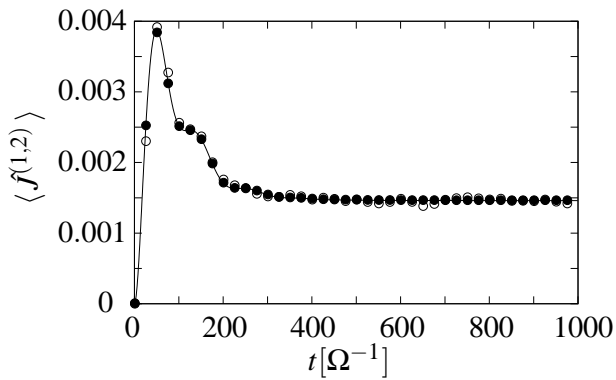}
\caption{Energy current $\langle\op{J}^{(1,2)}\rangle$ in a chain of $n=3$ weakly coupled spins, driven by heat baths of different temperature ($\beta_{\text L}=0.41,\,\beta_{\text R}=1.39,\,\lambda=\kappa=0.01,\,\Omega=1$). Solid line: computed on the basis of the Redfield master equation (\ref{twobathscenario}) by exact diagonalization in Liouville Space. Filled circles: weak internal coupling approximation (\ref{DDAqme}) (again exact diagonalization). Circles: Numerical solution of Eq. (\ref{DDAqme}) by means of a stochastic wave function simulation ($10^5$ realizations).}
\label{fig:weakcoupling}
\end{figure}

Figure~\ref{fig:weakcoupling} demonstrates the excellent match between the predictions of the Redfield master equation (\ref{rme}) and those of the Lindblad form (\ref{DDAqme}) that has been derived in this Section. Furthermore the results from a stochastic wave function simulation of Eq.~(\ref{DDAqme}) are depicted. We refer the reader to \cite{breuer,plenio-knight-1998} for details of the method. Investigations on larger chain models now become feasible and more detailed numerical results will be presented in a forthcoming paper.

Thus, the Eq.~(\ref{DDAqme}) contains both vital properties: 
On the one hand it is in Lindblad form, preserving all properties of the density matrix and being, furthermore,  well suited for all types of stochastical methods.
On the other hand it features a nonequilibrium steady state with an energy current flowing through the system.

The neglect of the interaction between spins in Eq.~(\ref{dda}) leads to the effect that the damping concerns the hypothetically uncoupled system only. 
All effects that arise from different internal coupling schemes will only show up in the coherent part of the QME.
However, the dynamics will not differ in the type of damping. One possible way to reintroduce the effects of the internal coupling might be a perturbational approach \cite{Kleinekathoefer}, but again positivity violation will arise most probably without further modifications.
Moreover we would like to point out that the neglect of the off-diagonal entries in Eq.~(\ref{coeff}) directly yields a Lindblad QME which is of the same type that has previously served for investigations on heat
transport in small quantum systems \cite{Michel2003} but hitherto lacked a microscopic derivation.

%
%
\section{Conclusions}
\label{CONCLU}

The purpose of this article was, on the one hand, to present a
model of an open system for which the standard microscopic
derivation of a Lindblad quantum master equation (QME) fails, in
the sense that it is inappropriate to model nonequilibrium steady
states when the system is subject to a temperature gradient
induced by baths of different temperature. 

On the other hand, we
have proposed an alternative derivation that does yield a Lindblad
QME and can properly describe energy currents in the stationary
state of weakly coupled spin-$\frac12$ Heisenberg chains. The
usefulness of a QME of Lindblad-form can not be overemphasized,
because it is the key link to efficient numerical investigations
by means of standard stochastic wave function methods.

Stress was laid on the fact that the secular approximation (SA)
performed on a quantum master equation for a non-degenerate,
one-dimensional system in a split bath scenario with temperature
gradient (see Fig.~\ref{fig:spinkette}) leads to a stationary state
which is diagonal in the energy eigenbasis of the system. Therefore, no energy
current is flowing from the hot towards the cold bath, a situation
which is highly unphysical. Hence, the use of a QME in the SA is
inappropriate for the description of this type of nonequilibrium
scenarios.

An open question remains whether the secular approximation
generally wipes out all local coupling aspects as considered in
this paper. Our studies have hitherto been restricted to
one-dimensional spin chains only. One primary future goal
therefore is to carry out further systematic investigations on the
question whether our findings are extendable to larger classes of
open systems and coupling models.

\begin{acknowledgments}
We thank G.~Mahler, M.~Hartmann and especially M.~Stollsteimer for fruitful discussions on the 
topic of nonequilibrium quantum master equations. Financial Support by the Deutsche
Forschungsgemeinschaft is gratefully acknowledged.
\end{acknowledgments}

\begin{thebibliography}{30}
\expandafter\ifx\csname natexlab\endcsname\relax\def\natexlab#1{#1}\fi
\expandafter\ifx\csname bibnamefont\endcsname\relax
  \def\bibnamefont#1{#1}\fi
\expandafter\ifx\csname bibfnamefont\endcsname\relax
  \def\bibfnamefont#1{#1}\fi
\expandafter\ifx\csname citenamefont\endcsname\relax
  \def\citenamefont#1{#1}\fi
\expandafter\ifx\csname url\endcsname\relax
  \def\url#1{\texttt{#1}}\fi
\expandafter\ifx\csname urlprefix\endcsname\relax\def\urlprefix{URL }\fi
\providecommand{\bibinfo}[2]{#2}
\providecommand{\eprint}[2][]{\url{#2}}

\bibitem[{\citenamefont{Lepri et~al.}(2003)\citenamefont{Lepri, Livi, and
  Politi}}]{Lepri2003}
\bibinfo{author}{\bibfnamefont{S.}~\bibnamefont{Lepri}},
  \bibinfo{author}{\bibfnamefont{R.}~\bibnamefont{Livi}}, \bibnamefont{and}
  \bibinfo{author}{\bibfnamefont{A.}~\bibnamefont{Politi}},
  \bibinfo{journal}{Phys. Rep.} \textbf{\bibinfo{volume}{377}},
  \bibinfo{pages}{1} (\bibinfo{year}{2003}).

\bibitem[{\citenamefont{Vollmer}(2002)}]{Vollmer2002}
\bibinfo{author}{\bibfnamefont{J.}~\bibnamefont{Vollmer}},
  \bibinfo{journal}{Phys. Rep.} \textbf{\bibinfo{volume}{372}},
  \bibinfo{pages}{131} (\bibinfo{year}{2002}).

\bibitem[{\citenamefont{Zotos et~al.}(1997)\citenamefont{Zotos, Naef, and
  Prelovsek}}]{zotos-1997}
\bibinfo{author}{\bibfnamefont{X.}~\bibnamefont{Zotos}},
  \bibinfo{author}{\bibfnamefont{F.}~\bibnamefont{Naef}}, \bibnamefont{and}
  \bibinfo{author}{\bibfnamefont{P.}~\bibnamefont{Prelovsek}},
  \bibinfo{journal}{Phys. Rev. B} \textbf{\bibinfo{volume}{55}},
  \bibinfo{pages}{11029} (\bibinfo{year}{1997}).

\bibitem[{\citenamefont{Jung et~al.}(2006)\citenamefont{Jung, Helmes, and
  Rosch}}]{Jung2006}
\bibinfo{author}{\bibfnamefont{P.}~\bibnamefont{Jung}},
  \bibinfo{author}{\bibfnamefont{R.}~\bibnamefont{Helmes}}, \bibnamefont{and}
  \bibinfo{author}{\bibfnamefont{A.}~\bibnamefont{Rosch}},
  \bibinfo{journal}{Phys. Rev. Lett.} \textbf{\bibinfo{volume}{96}},
  \bibinfo{pages}{067202} (\bibinfo{year}{2006}).

\bibitem[{\citenamefont{Michel et~al.}(2005)\citenamefont{Michel, Mahler, and
  Gemmer}}]{Michel2005}
\bibinfo{author}{\bibfnamefont{M.}~\bibnamefont{Michel}},
  \bibinfo{author}{\bibfnamefont{G.}~\bibnamefont{Mahler}}, \bibnamefont{and}
  \bibinfo{author}{\bibfnamefont{J.}~\bibnamefont{Gemmer}},
  \bibinfo{journal}{Phys. Rev. Lett.} \textbf{\bibinfo{volume}{95}},
  \bibinfo{pages}{180602} (\bibinfo{year}{2005}).

\bibitem[{\citenamefont{Kubo}(1957)}]{Kubo1957}
\bibinfo{author}{\bibfnamefont{R.}~\bibnamefont{Kubo}}, \bibinfo{journal}{J.
  Phys. Soc. Jpn.} \textbf{\bibinfo{volume}{12}}, \bibinfo{pages}{570}
  (\bibinfo{year}{1957}).

\bibitem[{\citenamefont{Heidrich-Meisner
  et~al.}(2002)\citenamefont{Heidrich-Meisner, Honecker, Cabra, and
  Brenig}}]{Heidrich2002}
\bibinfo{author}{\bibfnamefont{F.}~\bibnamefont{Heidrich-Meisner}},
  \bibinfo{author}{\bibfnamefont{A.}~\bibnamefont{Honecker}},
  \bibinfo{author}{\bibfnamefont{D.}~\bibnamefont{Cabra}}, \bibnamefont{and}
  \bibinfo{author}{\bibfnamefont{W.}~\bibnamefont{Brenig}},
  \bibinfo{journal}{Phys. Rev. B} \textbf{\bibinfo{volume}{66}},
  \bibinfo{pages}{140406} (\bibinfo{year}{2002}).

\bibitem[{\citenamefont{Michel et~al.}(2006)\citenamefont{Michel, Gemmer, and
  Mahler}}]{michel-2006-20}
\bibinfo{author}{\bibfnamefont{M.}~\bibnamefont{Michel}},
  \bibinfo{author}{\bibfnamefont{J.}~\bibnamefont{Gemmer}}, \bibnamefont{and}
  \bibinfo{author}{\bibfnamefont{G.}~\bibnamefont{Mahler}},
  \bibinfo{journal}{International Journal of Modern Physics B}
  \textbf{\bibinfo{volume}{20}}, \bibinfo{pages}{4855} (\bibinfo{year}{2006}).

\bibitem[{\citenamefont{Breuer and Petruccione}(2002)}]{breuer}
\bibinfo{author}{\bibfnamefont{H.-P.} \bibnamefont{Breuer}} \bibnamefont{and}
  \bibinfo{author}{\bibfnamefont{F.}~\bibnamefont{Petruccione}},
  \emph{\bibinfo{title}{The {T}heory of {O}pen {Q}uantum {S}ystems}}
  (\bibinfo{publisher}{Oxford University Press}, \bibinfo{year}{2002}).

\bibitem[{\citenamefont{Weiss}(1999)}]{Weiss1999}
\bibinfo{author}{\bibfnamefont{U.}~\bibnamefont{Weiss}},
  \emph{\bibinfo{title}{Quantum Dissipative Systems}}
  (\bibinfo{publisher}{World Scientific}, \bibinfo{address}{Singapore, New
  Jersey, London, Hong Kong}, \bibinfo{year}{1999}).

\bibitem[{\citenamefont{Scully and Zubairy}(1997)}]{Scully1997}
\bibinfo{author}{\bibfnamefont{M.~O.} \bibnamefont{Scully}} \bibnamefont{and}
  \bibinfo{author}{\bibfnamefont{M.~S.} \bibnamefont{Zubairy}},
  \emph{\bibinfo{title}{Quantum Optics}} (\bibinfo{publisher}{Cambridge
  University Press}, \bibinfo{address}{Cambridge}, \bibinfo{year}{1997}).

\bibitem[{\citenamefont{Saito et~al.}(2000)\citenamefont{Saito, Takesue, and
  Miyashita}}]{Saito2000}
\bibinfo{author}{\bibfnamefont{K.}~\bibnamefont{Saito}},
  \bibinfo{author}{\bibfnamefont{S.}~\bibnamefont{Takesue}}, \bibnamefont{and}
  \bibinfo{author}{\bibfnamefont{S.}~\bibnamefont{Miyashita}},
  \bibinfo{journal}{Phys. Rev. E} \textbf{\bibinfo{volume}{61}},
  \bibinfo{pages}{2397} (\bibinfo{year}{2000}).

\bibitem[{\citenamefont{Saito}(2003)}]{Saito2003}
\bibinfo{author}{\bibfnamefont{K.}~\bibnamefont{Saito}},
  \bibinfo{journal}{Europhys. Lett.} \textbf{\bibinfo{volume}{61}},
  \bibinfo{pages}{34} (\bibinfo{year}{2003}).

\bibitem[{\citenamefont{Michel et~al.}(2003)\citenamefont{Michel, Hartmann,
  Gemmer, and Mahler}}]{Michel2003}
\bibinfo{author}{\bibfnamefont{M.}~\bibnamefont{Michel}},
  \bibinfo{author}{\bibfnamefont{M.}~\bibnamefont{Hartmann}},
  \bibinfo{author}{\bibfnamefont{J.}~\bibnamefont{Gemmer}}, \bibnamefont{and}
  \bibinfo{author}{\bibfnamefont{G.}~\bibnamefont{Mahler}},
  \bibinfo{journal}{Euro. Phys. J. B} \textbf{\bibinfo{volume}{34}},
  \bibinfo{pages}{325} (\bibinfo{year}{2003}).

\bibitem[{\citenamefont{Gorrini et~al.}(1976)\citenamefont{Gorrini,
  Kossakowski, and Sudarshan}}]{Gorini1976}
\bibinfo{author}{\bibfnamefont{V.}~\bibnamefont{Gorrini}},
  \bibinfo{author}{\bibfnamefont{A.}~\bibnamefont{Kossakowski}},
  \bibnamefont{and}
  \bibinfo{author}{\bibfnamefont{E.}~\bibnamefont{Sudarshan}},
  \bibinfo{journal}{J. of Math. Phys.} \textbf{\bibinfo{volume}{17}},
  \bibinfo{pages}{821} (\bibinfo{year}{1976}).

\bibitem[{\citenamefont{Lindblad}(1976)}]{Lindblad1976}
\bibinfo{author}{\bibfnamefont{G.}~\bibnamefont{Lindblad}},
  \bibinfo{journal}{Commun. Math. Phys.} \textbf{\bibinfo{volume}{48}},
  \bibinfo{pages}{119} (\bibinfo{year}{1976}).

\bibitem[{\citenamefont{Plenio and Knight}(1998)}]{plenio-knight-1998}
\bibinfo{author}{\bibfnamefont{M.~B.} \bibnamefont{Plenio}} \bibnamefont{and}
  \bibinfo{author}{\bibfnamefont{P.~L.} \bibnamefont{Knight}},
  \bibinfo{journal}{Rev. Mod. Phys.} \textbf{\bibinfo{volume}{70}},
  \bibinfo{pages}{101} (\bibinfo{year}{1998}).

\bibitem[{\citenamefont{Davies}(1974)}]{Davies1974b}
\bibinfo{author}{\bibfnamefont{E.~B.} \bibnamefont{Davies}},
  \bibinfo{journal}{Commun. Math. Phys.} \textbf{\bibinfo{volume}{39}},
  \bibinfo{pages}{91} (\bibinfo{year}{1974}).

\bibitem[{\citenamefont{D\"umcke and Spohn}(1979)}]{Spohn}
\bibinfo{author}{\bibfnamefont{R.}~\bibnamefont{D\"umcke}} \bibnamefont{and}
  \bibinfo{author}{\bibfnamefont{H.}~\bibnamefont{Spohn}}, \bibinfo{journal}{Z.
  Phys. B} \textbf{\bibinfo{volume}{34}}, \bibinfo{pages}{419}
  (\bibinfo{year}{1979}).

\bibitem[{\citenamefont{Maniscalco et~al.}(2004)\citenamefont{Maniscalco,
  Intravaia, Piilo, and Messina}}]{maniscalco-2004-6}
\bibinfo{author}{\bibfnamefont{S.}~\bibnamefont{Maniscalco}},
  \bibinfo{author}{\bibfnamefont{F.}~\bibnamefont{Intravaia}},
  \bibinfo{author}{\bibfnamefont{J.}~\bibnamefont{Piilo}}, \bibnamefont{and}
  \bibinfo{author}{\bibfnamefont{A.}~\bibnamefont{Messina}},
  \bibinfo{journal}{J. Opt. B: Quantum Semiclass. Opt.}
  \textbf{\bibinfo{volume}{6}}, \bibinfo{pages}{98} (\bibinfo{year}{2004}).

\bibitem[{\citenamefont{{W.T. Pollard} et~al.}(1996)\citenamefont{{W.T.
  Pollard}, {A.K. Felts}, and R.A.Friesner}}]{pollard}
\bibinfo{author}{\bibnamefont{{W.T. Pollard}}},
  \bibinfo{author}{\bibnamefont{{A.K. Felts}}}, \bibnamefont{and}
  \bibinfo{author}{\bibnamefont{R.A.Friesner}}, \bibinfo{journal}{Advances in
  Chemical Physics} \textbf{\bibinfo{volume}{93}}, \bibinfo{pages}{77}
  (\bibinfo{year}{1996}).

\bibitem[{\citenamefont{{Volkhard May} and {Oliver Kuehn}}(2003)}]{may}
\bibinfo{author}{\bibnamefont{{Volkhard May}}} \bibnamefont{and}
  \bibinfo{author}{\bibnamefont{{Oliver Kuehn}}}, \emph{\bibinfo{title}{Charge
  and {E}nergy {T}ransfer {D}ynamics in {M}olecular {S}ystems}}
  (\bibinfo{publisher}{Wiley-VCH, Berlin}, \bibinfo{year}{2003}).

\bibitem[{\citenamefont{{S. Gnutzmann} and {F. Haake}}(1996)}]{Gnutzmann1996}
\bibinfo{author}{\bibnamefont{{S. Gnutzmann}}} \bibnamefont{and}
  \bibinfo{author}{\bibnamefont{{F. Haake}}}, \bibinfo{journal}{Zeitschrift
  f\"ur Physik B} \textbf{\bibinfo{volume}{101}}, \bibinfo{pages}{263}
  (\bibinfo{year}{1996}).

\bibitem[{\citenamefont{Suarez et~al.}(1992)\citenamefont{Suarez, Silbey, and
  Oppenheim}}]{Suarez1992}
\bibinfo{author}{\bibfnamefont{A.}~\bibnamefont{Suarez}},
  \bibinfo{author}{\bibfnamefont{R.}~\bibnamefont{Silbey}}, \bibnamefont{and}
  \bibinfo{author}{\bibfnamefont{I.}~\bibnamefont{Oppenheim}},
  \bibinfo{journal}{The Journal of Chemical Physics}
  \textbf{\bibinfo{volume}{97}}, \bibinfo{pages}{5101} (\bibinfo{year}{1992}).

\bibitem[{\citenamefont{Kubo et~al.}(1991)\citenamefont{Kubo, Toda, and
  Hashitsume}}]{Kubo1991}
\bibinfo{author}{\bibfnamefont{R.}~\bibnamefont{Kubo}},
  \bibinfo{author}{\bibfnamefont{M.}~\bibnamefont{Toda}}, \bibnamefont{and}
  \bibinfo{author}{\bibfnamefont{N.}~\bibnamefont{Hashitsume}},
  \emph{\bibinfo{title}{Statistical {P}hysics {II}: {N}onequilibrium
  {S}tatistical {M}echanics}}, no.~\bibinfo{number}{31} in
  \bibinfo{series}{Solid-State Sciences} (\bibinfo{publisher}{Springer},
  \bibinfo{address}{Berlin, Heidelberg, New-York}, \bibinfo{year}{1991}),
  \bibinfo{edition}{2nd} ed.

\bibitem[{\citenamefont{{H.-P. Breuer} et~al.}(1999)\citenamefont{{H.-P.
  Breuer}, {B. Kappler}, and {F. Petruccione}}}]{Breuer2}
\bibinfo{author}{\bibnamefont{{H.-P. Breuer}}},
  \bibinfo{author}{\bibnamefont{{B. Kappler}}}, \bibnamefont{and}
  \bibinfo{author}{\bibnamefont{{F. Petruccione}}}, \bibinfo{journal}{Physical
  Review A} \textbf{\bibinfo{volume}{59}}, \bibinfo{pages}{1633}
  (\bibinfo{year}{1999}).

\bibitem[{\citenamefont{Alicki and Lendi}(1986)}]{Alicki}
\bibinfo{author}{\bibfnamefont{R.}~\bibnamefont{Alicki}} \bibnamefont{and}
  \bibinfo{author}{\bibfnamefont{K.}~\bibnamefont{Lendi}},
  \emph{\bibinfo{title}{Quantum Dynamical Semigroups and Applications}}
  (\bibinfo{publisher}{Springer Verlag}, \bibinfo{year}{1986}).

\bibitem[{\citenamefont{Stollsteimer}(2006)}]{Stollsteimer2006}
\bibinfo{author}{\bibfnamefont{M.}~\bibnamefont{Stollsteimer}}, Ph.D. thesis,
  \bibinfo{school}{Universit\"at Stuttgart} (\bibinfo{year}{2006}).

\bibitem[{\citenamefont{Bl\"umel et~al.}(1991)\citenamefont{Bl\"umel,
  Buchleitner, Graham, Sirko, Smilansky, and Walther}}]{Blumel1991}
\bibinfo{author}{\bibfnamefont{R.}~\bibnamefont{Bl\"umel}},
  \bibinfo{author}{\bibfnamefont{A.}~\bibnamefont{Buchleitner}},
  \bibinfo{author}{\bibfnamefont{R.}~\bibnamefont{Graham}},
  \bibinfo{author}{\bibfnamefont{L.}~\bibnamefont{Sirko}},
  \bibinfo{author}{\bibfnamefont{U.}~\bibnamefont{Smilansky}},
  \bibnamefont{and} \bibinfo{author}{\bibfnamefont{H.}~\bibnamefont{Walther}},
  \bibinfo{journal}{Phys. Rev. A} \textbf{\bibinfo{volume}{44}},
  \bibinfo{pages}{4521 } (\bibinfo{year}{1991}).

\bibitem[{\citenamefont{Kleinekathoefer
  et~al.}(2001)\citenamefont{Kleinekathoefer, Kondov, and
  Schreiber}}]{Kleinekathoefer}
\bibinfo{author}{\bibfnamefont{U.}~\bibnamefont{Kleinekathoefer}},
  \bibinfo{author}{\bibfnamefont{I.}~\bibnamefont{Kondov}}, \bibnamefont{and}
  \bibinfo{author}{\bibfnamefont{M.}~\bibnamefont{Schreiber}},
  \bibinfo{journal}{Chem.Phys.} \textbf{\bibinfo{volume}{268}},
  \bibinfo{pages}{121} (\bibinfo{year}{2001}).

\end{thebibliography}

\end{document}